\documentclass[aps,showpacs,amsmath,amssymbd,dvips,showkeys,preprint]{revtex4}

\usepackage{graphicx}
\usepackage{color}

\def \be  {\begin{equation}}
\def \ee  {\end{equation}}
\def \ee  {\end{equation}}
\def \bea {\begin{eqnarray}}
\def \eea {\end{eqnarray}}

\newcommand{\nn}{\nonumber}

\begin{document}

\preprint{ECTP-2015-10}    
\preprint{WLCAPP-2015-10}
\vspace*{.5cm}

\title{Perturbative Instability of Cosmology from Quantum Potential}

\author{Abdel~Nasser~Tawfik\footnote{http://atawfik.net/}}
\affiliation{Egyptian Center for Theoretical Physics (ECTP), Modern University for Technology and Information (MTI), 11571 Cairo, Egypt}
\affiliation{World Laboratory for Cosmology And Particle Physics (WLCAPP), Cairo, Egypt}

\author{Abdel~Magied~Diab}
\affiliation{Egyptian Center for Theoretical Physics (ECTP), Modern University for Technology and Information (MTI), 11571 Cairo, Egypt}
\affiliation{World Laboratory for Cosmology And Particle Physics (WLCAPP), Cairo, Egypt}

\author{Eiman Abou El Dahab}
\affiliation{Faculty of Computer Science, Modern University for Technology and Information (MTI), 11571 Cairo, Egypt}
\affiliation{World Laboratory for Cosmology And Particle Physics (WLCAPP), Cairo, Egypt}

\author{Tiberiu Harko}
\affiliation{Department of Mathematics, University College London,
Gower Street, London WC1E 6BT, United Kingdom}

\date{\today}

\begin{abstract}
Apart from its debatable correctness, we examine the perturbative stability of the recently proposed cosmology from quantum potential. We find that the proposed quantum corrections invoke additional parameters which apparently introduce perturbative instability to the  Universe.
\end{abstract}


\maketitle

\section{Introduction}
\label{sec:intr}

The standard model for cosmology assumes a high degree of precision with regards to the spatially homogeneous and isotropic structure of our Universe \cite{reffrr1}, which is well described by the Friedmann-Lemaitre-Robertson-Walker (FLRW) metric. This leads to restrictions on the possible geometric topologies of the large-scale structure of the Universe, i.e. either closed, open, or flat FLRW spaces. Nevertheless, any proposed world model should be able to describe the expanding Universe and simultaneously show that the resulting Universe is stable, as is observed. 

In general, there are different kinds of stability criteria: a) a minimum energy level ensuring that the physical system does not collapse into negative-energy levels, b) nothing is allowed to be created out of nothing, ensuring physical conservations, and c) arbitrary small perturbations should not drive the system out of equilibrium. Studying the stability of the Einstein universe dates back to the 1930's; for instance, Eddington-Lemaitre's picture of the {\it ''primordial atom''} and found instability against spatially homogeneous and isotropic perturbations \cite{reffrr2}.  

The structural stability - which has been well known since the early stages of Einstein cosmology - has led numerous successes. Structural stability should also be applied to all subsystems \cite{reffrr3}. Alternatively, perturbative stability is a powerful tool based on understanding the physics of the perturbation propagation. The introduction of a discrete perturbation at an arbitrary point is then followed by analyzing its effects over time.

The stability criterion  is fulfilled if the added perturbation changes at a later time \cite{Dobado:1995, Kao:1991,Kao:2001a, Kao:2001b}. Such a stability analysis for nonredundant field equations in a Bianchi type I universe has been performed in the isotropic limit \cite{Kao:2001b} and for anisotropic brane cosmology \cite{Kao:2002s}.
\begin{itemize}
\item For a Bianchi type I isotropic brane cosmology \cite{Kao:2001b}, it was shown that any unstable mode of the isotropic perturbation with respect to a de Sitter background is also unstable with respect to anisotropic perturbations. This type of world model is stable against any anisotropic perturbation for a perfect fluid or a dilaton field  \cite{Kao:2002s}.
\item  In the large-time limit and independent of the different types of background matter, the anisotropic expansion of the anisotropic brane cosmology is dynamically smeared out \cite{Kao:2002s}. In addition, the stability analysis \cite{Harko:2001s, Kao:2001cx, Maroto:1995tr, Maroto:1997se, Kao:200fw} indicates that all such models are stable against any anisotropic perturbation. The perturbative (in)stablity is conditioned by the existence of a mode in the plane-wave equation, where $\gamma_+>0$ (unstable) or $\gamma_-<0$ (stable), respectively. 
\end{itemize}

The impacts of dark matter \cite{Ellis:2003,Primack:2003}, dark energy \cite{Sahni:2000,Peebles:2003s,Alam:2004pl,Copeland:2006,Padmanabhan:2003es,Straumann:2006po}, and the cosmological constant \cite{Weinberg:1989,Carroll:2001} on reliable (stable) world models have been evaluated.  Nevertheless, the of a static, closed and singularity-free universe - known as the "{\it emergence of cosmic space}" has been proposed, recently - was recently proposed 
 \cite{Padm12,Mulryne:2005s,Guendelman:2011,Wu:2010fg,Ellis:2004sg,Bandyopadhyay:2007ss,Bandyopadhyay:2007fh}. Of particular relevance is the instability of the static Einstein universe, especially for infinitely long times in presence of quantum fluctuations. Furthermore, it was found that the Einstein static Universe is unstable with respect to small radial perturbations \cite{Audrey,Mulryne:2005s,Guendelman:2011,Wu:2010fg}. Even if such models are perfectly fine-tuned to describe the early stages of the Universe, the quantum fluctuations among others would generate inflation or even entirely collapse it at infinite time (infinite age!) \cite{Mulryne:2005s,Guendelman:2011,Wu:2010fg}. 
 
The avoidance of an initial singularity \cite{Geroch:1966,Penrose:1970,Hawking:1973} - as explored in the idea of am {\it emergent universe} - motivated the introduction of quantum corrections \cite{Ali:2015,Lashin:2015}. It was argued that replacing the classical geodesic with Bohmian trajectories leads to quantum corrections to the Raychaudhuri equations \cite{Ali:2015}. Then, by deriving the Friedmann equations, the authors claimed that their corrections contain a correct estimation for the cosmological constant and the so-called radiation term. They have interpreted that the latter evades the big-bang singularity and determines an infinite age for the universe. Instead of criticizing the correctness of this approach as was done in Ref. \cite{Lashin:2015} - where it was argued that both conclusions are simply  wrong - we go another way. Instead of proposing radical corrections, we analyze the perturbative stability of both versions.

The present work is organized as follows. In Sec. \ref{sec_standard}, we implement the perturbative stability in a FLRW universe. The perturbative stability of FLRW cosmology from a quantum potential - which was proposed in Ref. \cite{Ali:2015} and criticized in Ref.  \cite{Lashin:2015} - shall be elaborated in Sec. \ref{Ray:Quan}. Section \ref{conc} is devoted to the discussion and final conclusions.

\section{Perturbative Cosmological Stability}
\label{stablUNI}

\subsection{FLRW cosmology}
\label{sec_standard}

The FLRW metric can be given as \cite{Kao:1991}
\bea
ds^2 = - b^2(t)\, dt^2 + a^2(t)\, \left( \frac{dr^2}{1-\kappa r^2} + r^2\, d\Omega\right),  
\eea
where $\kappa$ is the curvature constant $0,\,\pm 1$ stand for a flat, closed, or open universe, respectively, $a(t)$ is the scale factor, and $b(t)$ is the lapse function. For a perfect fluid, the energy-momentum tensor reads
\bea
T_{\mu \nu} &=& (\rho + p) \, u_{\mu}\, u_{\nu} + g_{ab}\, p,
\eea
where $u_\mu$ is the four-velocity field, $\mu$ and $\nu$ run over $0, \cdots, 3$, $\rho$ is the comoving energy density, and $p$ is the pressure. The energy-momentum conservation condition $D_\mu \, T^{\mu \nu}=0$ is apparently equivalent to the time evolution of the energy density which defines the continuity equation,
\bea
\dot{\rho} &=& -3 \rho (1+w) H.
\eea
At finite cosmological constant, the second form of the Friedmann equation which is also known as the Raychaudhuri equation gives 
\bea
\dot{H}= -\frac{3}{2} \,(1+w)\, H^2 \,+\, \frac{\Lambda \,c^2}{3} ,   \label{FRW_stand}
\eea
where the Hubble parameter $H=\dot{a}/a$, and $\omega =p/\rho$ is the equation of state (EoS). The cosmological constant ($\Lambda$) has dimensions of (length)$^{-2}$, and the approximate value  $\sim 10^{-52}\,$ m$^{-2}$.

By applying an infinitesimal perturbation to the Hubble parameter $H = \bar{H} + \delta\, H(t,x)$, the time evolution of $H$, the energy density, the pressure, and the equation of state, respectively, can be given as
\bea 
\dot{H} = \bar{\dot{H}}\,+\, \delta\, \dot{H}(t,x), \qquad
\rho = \bar{\rho} + \delta\, \rho(t,x), \qquad
p = \bar{p} + \delta\, p(t,x), \qquad 
\omega = \bar{\omega} + \delta\, \omega(t,x),
\label{pert}
\eea
where bar donates the spatial average. First, let us assume that $\alpha=-\frac{3}{2} (1+\omega)$,  
\bea
\alpha + \delta \alpha &=& -\frac{3}{2} (1+\bar{\omega} + \delta\, \omega).
\eea 
Then, by eliminating the higher orders, the frictional perturbation ($\delta \equiv  \delta \rho /\rho$) leads to  
\begin{eqnarray}
\delta \alpha &=&  -\frac{3}{2} \delta \omega = -\frac{3}{2} (\omega+\;\delta\omega\;- \omega) =  -\frac{3}{2} \Big[ \frac{p+\delta p}{\rho+\delta} -\frac{p}{\rho} \Big] =-\frac{3}{2}\frac{p}{\rho} \left[ \frac{1+ \left(\frac{\delta p}{p}\right)}{1+ \left(\frac{\delta \rho}{\rho}\right)} -1 \right] \nonumber \\  &=&  -\frac{3}{2}\frac{p}{\rho} \left[ \left( 1+ \frac{\delta p}{p}\right) \left(1+ \frac{\delta \rho}{\rho}\right) -1 \right] =  \frac{3}{2}\frac{p}{\rho} \frac{\delta \rho}{\rho} = \frac{3}{2} \omega \delta, \label{Exp1}
\end{eqnarray}
which obviously means that $\delta \alpha = -3/2 \delta\, \omega$. For an infinitesimal perturbation $\delta \, \bar{p} \ll\, \delta \,\rho \ll\, 1$, the first-order perturbations in the FLRW Raychaudhuri equation and the continuity equation, respectively, are  given as 
\bea
 \delta \dot{H} &=& -3 (1+\omega) H \delta H + \frac{3}{2} H^2 \omega \delta +\frac{c^2}{3}  \delta \Lambda, \label{delH} \\
 \delta \dot{\rho} &=& -3 (1+ \omega) \rho \delta H - 3H\delta \rho. \label{delrho}
\eea
Let $\dot{\delta} = \partial/\partial t \left(\delta \rho/\rho\right)$; then, according to Eq. (\ref{Exp1}) one obtains $\omega \delta \equiv - \delta \omega$,  and Eq. (\ref{delrho}) can be written as 
\bea
\dot{\delta} &=& -3 (1+\omega)  \delta H -3H \omega \delta. \label{delrho2} 
\eea
From the coupling between Eqs. (\ref{delrho2}) and (\ref{delH}), it is straightforward to determine the second time derivative  of Eq. (\ref{delH}) with a finite inhomogeneous $\Lambda$ term,
\bea
A\, \ddot{\delta H} + B\, \dot{\delta H} + C \, \delta H  = \frac{c^2}{3} \left( D\, \dot{\delta \Lambda} \,+\,E \, \delta \Lambda \right), \label{sol1}
\eea
where $A=D=1$, $B=-3 (2+3\omega) H$, $C= (-9/2)(1+ 5\omega) H^2$ and $E =-3H$. The general solution of the inhomogeneous ordinary differential equation (\ref{sol1}) reads
\bea
\delta H (t) &=& \beta_1 \exp{\left[ \gamma_+ \,t\right]}\,+\, \beta_2 \exp{\left[ \gamma_- \,t\right]} +  \nn \\ && \,  \exp{\left[ \gamma_- \,t\right]} \cdot \int_ 1 ^t \Bigg( \frac{\exp{\left[ (-3H(2+3\omega)+\gamma_+) \,K  \right]} \,  c^2 \, \sqrt{3H (6+\omega + 9 \omega^2)} \, \delta \Lambda(K) }{3H (6+\omega + 9 \omega^2)} \nn \\ &&\,-\, \frac{\exp{\left[ (-3H(2+3\omega)+\gamma_+)\, K  \right]} \,  c^2 \, \sqrt{3H (6+\omega + 9 \omega^2)} \, \dot{\delta} \Lambda(K) }{3H^2 (6+\omega + 9 \omega^2)} \Bigg) \, dK \nn \\ && \,+\, \exp{\left[ \gamma_+ \,t\right]} \cdot \int_ 1 ^t  \Bigg( \frac{\exp{\left[ (-3H(2+3\omega)+\gamma_-) \, K  \right]} \,  c^2 \, \sqrt{3H (6+\omega + 9 \omega^2)} \, \delta \Lambda(K) }{3H (6+\omega + 9 \omega^2)} \nn \\ &&\,-\, \frac{\exp{\left[ (-3H(2+3\omega)+\gamma_-) \, K  \right]} \,  c^2 \, \sqrt{3H (6+\omega + 9 \omega^2)} \, \dot{\delta} \Lambda(K) }{3H^2 (6+\omega + 9 \omega^2)} \Bigg) \, dK,  \label{solnA2}
\eea 
where $\gamma_\pm$ shall be given in Eq. (\ref{simpSolA}). If $\Lambda$ is finite but its time derivative vanishes, then the second and fourth integrals should be removed. 

If $\Lambda$ terms vanish, Eq. (\ref{solnA2}) becomes homogeneous and can be solved as
\bea
\delta H (t) &=& \beta_1 \exp{\left[ \gamma_+ \,t\right]}\,+\, \beta_2 \exp{\left[ \gamma_- \,t\right]}, \label{solnA}
\eea 
where $\gamma_{\pm} = -B\pm \sqrt{B^2\,-4\,A\,C}/2A,$ and the parameters $\beta_1$ and  $\beta_2$ can be determined from initial perturbations. The zeroth-order perturbation gives exactly the field equation for the background field, i.e., $H\equiv H_0$. The exponent term can be simplified as,
\bea
\gamma_{\pm} = \frac{3}{2} \left[ (2+3\omega) \pm \sqrt{6+22\omega + 9\omega^2  }\right] H_0. \label{simpSolA}
\eea

It is apparent that even the inhomogeneous ordinary differential equation does not help in optimizing the perturbative stability of the standard FLRW Universe. From Eq. (\ref{simpSolA}), it is obvious that the first term on the right-hand side is always positive as $\omega$ and $H_0$ are positive quantities. This part apparently refers to a stable mode and isotropic perturbation. The second term can be negative,  referring to unstable modes and anisotropic perturbation. Accordingly, the stability conditions can be determined from Eq. (\ref{simpSolA}). It is stable at $\gamma_-<0$ and unstable at $\gamma_+>0$. 
\begin{itemize}
\item
The square root is less than the first two terms, $3 (2+3\omega)$. Then, the solution is apparently identical to the stability equation for the existence of an inflationary phase era of the de Sitter solution \cite{Dobado:1995,Kao:2001a,Kao:2001b}. 
\item
Occasionally, the square root might possess instability modes, i.e. $\gamma_+ >,0$. In this case, despite the fact that the inflationary era shall come to an end once such an unstable mode takes place, the latter  likely sharpens the stability of the isotropic space \cite{Kao:2001a,Kao:2001b}. It has been concluded that even if such an unstable mode for the de Sitter perturbation were to happen, it will be unstable against the anisotropic perturbation.
\end{itemize}

Thus, we conclude that for a standard FLRW universe, the stable modes are characterized by positive a EoS, $~\omega\geq 0$, especially in the matter-/radiation-dominated eras. For negative dark energy, the EoS might be negative, $-1<\omega<-1/3$, leading to instability with respect to a small perturbation. The unstable modes exist for negative $\omega$ without including a cosmological constant. This means that the matter-/radiation-dominated eras are appropriated ranges to interpret the stability with respect to arbitrarily small perturbations of the FLRW universe. For a dark energy equation of state, i.e., negative $\omega$, the universe is likely unstable against the anisotropic perturbation.

\subsection{FLRW cosmology with quantum corrections}
\label{Ray:Quan}

In this section to investigate whether or not the FLRW cosmology remains stable in quantum theory. By replacing the classical geodesic with quantum trajectories \cite{Das:2014}, the Raychaudhuri equations get quantum corrections \cite{Ali:2015},  
\bea
\dot{H}= -\frac{3}{2} \,(1+w)\, H^2 - \frac{6 \epsilon_1 \hbar^2}{m^2}\, (1+w) \left[ \,6(1+w)^2 -\, \frac{81}{2}(1+w)+\, 18\, \right] H^4. \label{FRW_quanta1}
\eea
This was criticized and accordingly considerable corrections have been proposed \cite{Lashin:2015},
\bea
\dot{H}= -\frac{3}{2} \,(1+w)\, H^2 - \frac{9 \hbar^2}{4 m^2 c^4} \epsilon (1-9 \omega^2) (1+\omega) H^4, \label{FRW_quanta2}
\eea
where the arbitrary constants $\epsilon_1$ and $\epsilon$ might differ from each other.

As discussed in the Introduction, we do not intend to comment on the incorrectness of Eq. (\ref{FRW_quanta1}) \cite{Ali:2015} and/or approve the proposal of  Eq. (\ref{FRW_quanta2}) \cite{Lashin:2015}. The present work is merely devoted to checking the perturbative (in)stability. Accordingly, one can judge whether this proposal or the other (or both) are physically (ir)relevant. The authors of Ref. \cite{Ali:2015} did not want to give details about the numerical factor in front of $\hbar^2$ and $m^2$ \cite{Das:2014}. Without clear scientific argumentation, they categorically rejected the proposed corrections of Eq. (\ref{FRW_quanta2}) \cite{Lashin:2015}! Our goal is the introduction of a systematic study for the perturbative stability of both equations by evaluating their (un)stable modes. 

Let us assume that the correction part in Eq. (\ref{FRW_quanta1}) is given as 
\bea
\xi &=& - \frac{6 \epsilon_1 \hbar^2}{m^2}\, (1+w) \left[ \,6(1+w)^2 -\, \frac{81}{2}(1+w)+\, 18\, \right]. \label{eq:qcorrALI}
\eea
By applying perturbations as in Eq. (\ref{pert}), the first-order perturbation of $\omega$ reads
\bea
\delta \xi = - 270 \frac{\epsilon_1 \hbar^2}{m^2} \, \omega \delta, \label{dxi}
\eea
where $\omega \delta \equiv - \delta \omega$. Accordingly, the perturbation of Eq. (\ref{FRW_quanta1})  becomes 
\bea
\dot{\delta H} & = & (2 \alpha H + 4 \xi H^3)\, \delta H \,+\, \left(\frac{3}{2} H^2  - \frac{270 \epsilon_1 \hbar^2}{m^2} H^4 \right)  \omega \delta. \label{pert03}
\eea

By using Eqs. (\ref{delrho2}) and (\ref{pert03}) - which can be reexpressed, respectively, as
\bea
\dot{\delta H} & = & \lambda_1 \,\delta H \,+\,  \lambda_2 \,\delta, \label{pert04} \\
\dot{\delta} & = & \lambda_3\, \delta H \,+\,  \lambda_4\, \delta, \label{deldot0}
\eea
where $\lambda_1 = 2 \alpha H + 4 \xi H^3$, $\lambda_2 =\left(3\, H^2/2  - (270 \epsilon_1 \hbar^2/m^2) H^4 \right)\omega$,  $\lambda_3 =-3(1+\omega)$ and $\lambda_4 = -3 H \omega$ - it is straightforward to determine the second time derivative of Eq. (\ref{pert04}) and by using Eq. (\ref{deldot0}) we obtain
\bea
{\cal A}_1\, \ddot{\delta H} + {\cal B}_1\, \dot{\delta H} + {\cal C}_1 \, \delta H = 0, \label{sol01}
\eea
where ${\cal A}_1=1$, ${\cal B}_1= - (\lambda_1 + \dot{\lambda_2}/\lambda_2 + \lambda_4)$  and ${\cal C}_1= - (\dot{\lambda_1} -\lambda_1\, \dot{\lambda_2}/\lambda_2  + \lambda_2 \lambda_3 - \lambda_1 \lambda_4)$. Moreover, the constants of Eq. (\ref{sol01}) can be simplified as follows:
\bea
{\cal B}_1&=& -\left[ \lambda_1 + \frac{\dot{\lambda_2}}{\lambda_2} + \lambda_4 \right]=
-\left[  2\alpha H + 4 \xi H^3 - 3H \omega + \frac{2(\alpha H + \xi H^3)\, (1 - \frac{4}{3} \frac{270 \epsilon_1 \hbar^2}{m^2}\,  H^2)}{ (1 - \frac{2}{3} \frac{270 \epsilon_1 \hbar^2}{m^2}\,  H^2)} \right], \hspace*{8mm}
\eea
where $\dot{H}/H =\alpha H + \xi H^3$. Also, the third term becomes
\bea
{\cal C}_1= -\left[\dot{\lambda_1} -\lambda_1\,\frac{\dot{\lambda_2}}{\lambda_2}  + \lambda_2 \lambda_3 - \lambda_1 \lambda_4 \right] &=& 2 (\alpha + 6 \xi H^2) (\alpha H^2 + \xi H^4) \\ \nn &-&
\frac{2(\alpha H + \xi H^3) \, (2 \alpha H+ 4\xi H^3) \, (1 - \frac{4}{3} \frac{270 \epsilon_1 \hbar^2}{m^2}\,  H^2)}{ (1 - \frac{2}{3} \frac{270 \epsilon_1 \hbar^2}{m^2}\,  H^2)} \\ \nn &-& 3 (1+\omega)  \left(\frac{3}{2} H^2  - \frac{270 \epsilon_1 \hbar^2}{m^2} H^4 \right)\omega + 3 H\omega \, (2 \alpha H+ 4\xi H^3).
\eea 

Again, the general solution of Eq. (\ref{sol01}) is
\bea
\delta H (t) &=& \beta_3 \exp{\left[ \gamma_+ \,t\right]}\,+\, \beta_4 \exp{\left[ \gamma_- \,t\right]}, \label{solnA}
\eea 
where the parameters $\beta_3$ and $\beta_4$ can be determined from the initial perturbations. The cosmological constant ($\Lambda$) was omitted as Ref. \cite{Ali:2015} claimed that the quantum correction, (\ref{eq:qcorrALI}) includes $\Lambda$ and moreover gives an exact estimation for it. Then, the modes $\gamma_ \pm$ are given as,
\bea
\gamma_ \pm & = & \left. \frac{3}{2} H_0 \Bigg( (2+ 3 \omega) -  3 \frac{270 \epsilon_1 \hbar^2}{m^2}\,  H_0 ^2 (11+30 \omega)  \right.  \\ \nn && \pm \, \, \Bigg[  6 +  \omega (22+17\omega) + \frac{2}{3} \frac{270 \epsilon_1 \hbar^2}{m^2}\, (1+\omega) (-209+\omega (-792 + \omega (-665+ 156 \omega))) \,  H^2 + \\ \nn && 17 \left(\frac{270 \epsilon_1 \hbar^2}{m^2} \right)^2 (11+\omega(30+ \omega(15-4\omega)))^2 \, H^4\, \, \Bigg)
\Bigg]^{1/2}
\eea
The first four terms are always positive. Thus, the square root defines (un)stable modes.

Also, for the corrected version (\ref{FRW_quanta2}) we follow the same procedure. First, we assume that the correction term of Eq. (\ref{FRW_quanta2}) is given as
\bea
\zeta &=& - \frac{9 \hbar^2}{4 m^2 c^4} \epsilon (1-9 \omega^2) (1+\omega),
\eea
The first-order perturbation in $\omega$ leads to 
\bea
\delta \, \zeta &=&+\, \frac{9 \hbar^2}{4 m^2 c^4}\, \epsilon\,  \omega \delta.
\eea
As given in Eq. (\ref{pert04}), the perturbation of Eq. (\ref{FRW_quanta2}) is
\bea
\dot{\delta H} & = & \lambda_5 \, \delta H \,+\, \lambda_6   \delta, \label{pert05}
\eea
where $\lambda_5 =  2 \alpha H + 4 \zeta H^3$  and $\lambda_6=  \left[(3/2) H^2  + (9 \epsilon \hbar^2)/(4 \,m^2 c^4) H^4\right] \,\omega $, By using Eq. (\ref{deldot0}), and as in Eq. (\ref{sol01}), we can determine the second time derivative of  Eq. (\ref{pert05}) as done before but with ${\cal A}_2=1$, ${\cal B}_2= - (\lambda_5 + \dot{\lambda_6}/\lambda_6 + \lambda_4)$  and ${\cal C}_2= - (\dot{\lambda_5} -\lambda_5\, \dot{\lambda_6}/\lambda_6  + \lambda_6 \lambda_3 - \lambda_5 \lambda_4)$
\bea
{\cal B}_2&=& -\left[ \lambda_5 + \frac{\dot{\lambda_6}}{\lambda_6} + \lambda_4 \right]=
-\left[  2\alpha H + 4 \zeta H^3 - 3H \omega + \frac{2(\alpha H + \zeta H^3)\, (1 + \frac{4}{3}\frac{9 \hbar^2}{4 m^2 c^4}\, \epsilon\,  \,  H^2)}{ (1 + \frac{2}{3} \frac{9 \hbar^2}{4 m^2 c^4}\, \epsilon\,  \,  H^2)} \right],  \hspace*{5mm} \\
 {\cal C}_2 &=& -\left[\dot{\lambda_5} -\lambda_5\,\frac{\dot{\lambda_6}}{\lambda_6}  + \lambda_6 \lambda_3 - \lambda_5 \lambda_4 \right] = 2 (\alpha + 6 \zeta H^2) (\alpha H^2 + \zeta H^4) \nn \\ &-&
\frac{2(\alpha H + \zeta H^3) \, (2 \alpha H+ 4\zeta H^3) \, (1 + \frac{4}{3} \frac{9 \hbar^2}{4 m^2 c^4}\, \epsilon\,  \,  H^2)}{ (1 + \frac{2}{3} \frac{9 \hbar^2}{4 m^2 c^4}\, \epsilon\,  \,  H^2)} \nn \\  &-& 3 (1+\omega)  \left(\frac{3}{2} H^2  +\frac{9 \hbar^2}{4 m^2 c^4}\, \epsilon\,  H^4 \right)\omega + 3 H\omega \, (2 \alpha H+ 4\zeta H^3).
\eea 

As in Eqs. (\ref{solnA2})  and (\ref{sol01}), the general solution reads 
 \bea
\delta H (t) &=& \beta_5 \exp{\left[ \gamma_+ \,t\right]}\,+\, \beta_6 \exp{\left[ \gamma_- \,t\right]}, \label{solnA}
\eea 
where the parameters $\beta_5$, and $\beta_6$ can determined by the initial perturbations. The modes $\gamma_ \pm$ are given as,
\bea
\gamma_ \pm & = & \left. \frac{3}{2} H_0 \Bigg( (2+ 3 \omega) -  2\frac{9 \hbar^2}{4 m^2 c^4}\,  H_0 ^2 (1+ \omega)  \right.  \\ \nn && \pm \, \,  \Bigg[  6 +\omega (22+17 \omega) +\frac{ \hbar^2}{m^2 c^4} \, (1+\omega) (19 + \omega (42+\omega (19+39\omega))) H^2 _0 +\\ \nn &&  \frac{68}{9} \left(\frac{9 \hbar^2}{4 m^2 c^4} \right)^2(1+2\omega)(1+\omega + \omega^2 + \omega^3)   H_0 ^4\Bigg]^{1/2} \, \, \Bigg).
\eea
Also, here the first four terms are always positive. The (un)stable modes are given by the last term, i.e., the square root.

\section{Discussion and Conclusions}
\label{conc}

The numerical estimation for (un)stable modes plays an essential role in determining the Universe (in)stability. The analysis of the perturbative (in)stability of the standard FLRW universe and that from the quantum corrections to the Raychaudhuri equations is strongly dependent on the choices for the parameters. We assumes that the parameters $\epsilon$ or $\epsilon_1$, $\hbar$, and the mass $m$ have the values $1/6$, $4.135\times10^{-15}$ eV s, and $\sim 10^{-32}~$eV/$c^2$, respectively \cite{Ali:2015}. In Ref. \cite{Lashin:2015}, the uncertainties in the physical quantities have been discussed. The authors of Ref. \cite{Ali:2015} did not want to even mention how they have evaluated their parameters!

For various equations of state - including dark energy and a cosmological constant (negative $\omega$), matter-/radiation-dominated eras (positive $\omega$) and an additional one characterized by $\omega=5.27$ - the (un)stable modes ($\gamma_\pm$) are ''not defined''  for cosmology with a quantum potential [Eq. (\ref{FRW_quanta1})  and  (\ref{FRW_quanta2})], i.e., imaginary values (nonphysical solutions). In the case of physical solutions, we can estimate the modes of Eqs. (\ref{FRW_quanta1}) and (\ref{FRW_quanta2}), where their signs directly point to a (un)stable universe. We conclude that the resulting modes are simply badly unstable and the solutions are nonphysical. 

In inflation and accelerating expansion, the equation of state is characterized by a negative $\omega$. According to observations for the {\it unseen} Universe \cite{Hogan:2007a}, $\omega\approx -1$. For $\omega > -1$, the dark energy density slowly decreases as the universe expands, but it increases for $\omega < -1$.  At $\omega=-1$, both equations become strongly dependent on the choice of the parameters $\epsilon$ or $\epsilon_1$, $\hbar$, and $m$. It worthwhile to recall that the standard FLRW universe at vanishing $\Lambda$ and $\omega=-1$ is unstable against a small perturbation. By replacing classical trajectories in such an Einstein static state (which is nothing but an {\it emergent Universe}) with the quantum ones, the instability becomes noteworthy. We conclude that the quantum corrections add additional parameters ($\epsilon$ or $\epsilon_1$, $\hbar$, and $m$) which apparently heighten the perturbative instability of our Universe.


\end{document}